\documentstyle[12pt, epsf, aasms4]{article} 

\begin{document} 

\noindent
{\it To appear in the September 21, 2000 issue of Nature }

\title{
The Accelerations of Stars Orbiting  \\
the Milky Way's Central Black Hole 
}
\author{A. M. Ghez$^*$, M. Morris$^*$, E. E. Becklin$^*$, A. Tanner$^*$, 
\& T. Kremenek$^*$}
\affil{$^*$Dept. of Physics and Astronomy, UCLA, Los Angeles, CA 90095-1562}

{\bf 
Recent measurements$^{1-4}$,
of the velocities of stars near the center of the
Milky Way have provided the strongest evidence for the presence of a
supermassive black hole in a galaxy$^5$, but the observational uncertainties
poorly constrain many of the properties of the black hole.
Determining the accelerations of stars in their orbits around
the center provides much more precise information about the
position and mass of the black hole.
Here we report measurements of the accelerations for three stars located
$\sim$ 0.005 pc (projected on the sky) from the central
radio source Sagittarius A* (Sgr A*); 
these accelerations are comparable to those experienced by the Earth as it 
orbits the Sun.  
These data increase
the inferred minimum mass density in the central region of the Galaxy
by an order of magnitude relative to previous results
and localized the dark mass to within
0.05 $\pm$ 0.04 arcsec of the nominal position of Sgr A*.
In addition, the orbital period of one of the
observed stars could be as short as 15 years, allowing us the
opportunity in the near future to observe an entire period. }

In 1995, we initiated a program of high resolution 2.2 $\mu m$ (K band)
imaging of the inner 5 arcsec $\times$ 5 arcsec of the Galaxy's central 
stellar cluster with the W. M. Keck 10-m telescope on Mauna Kea, Hawaii
(1 arcsec = 0.04 pc at the distance to the Galactic Center, 8 kpc, ref. 6).
From each observation, 
several thousand short exposure ($t_{exp}$=0.137 sec) frames 
were collected, using 
the facility near-infrared camera, NIRC$^{7,8}$,
and combined to produce a
final diffraction-limited image, having an angular resolution of 
0.05 arcsec.
Between 1995 and 1997, images were obtained once a 
year with the aim of detecting the stars' velocities
in the plane of the sky.  The results from these 
measurements
are detailed in ref. 4; in summary, two-dimensional velocities 
were measured for 90 stars with simple linear fits to the positions
as a function of time.  These velocities, which reach up to
1400 km/sec, implied
the existence of a $2.6 \times 10^6 {\mathrm M}_{\odot}$ black hole 
coincident ($\pm$0.1 arcsec) with the nominal location of 
Sgr A* (ref. 9),  the unusual radio source$^{10,11,12}$ 
long believed to be the counterpart of the putative black hole. 

The new observations presented here were obtained   
several times a year from 1997 to 1999 
to improve the sensitivity to accelerations.
With nine independent measurements,
we now fit the positions of 
stars as a function of time with second order polynomials (see Fig. 1).
The resulting velocity uncertainties are reduced by a factor of 3 
compared to our earlier work, primarily as a result of the increased
time baseline, and, in the central square arcsecond, 
by a factor of 6 that presented in ref. 3,
due to our higher angular resolution.
Among the 90 stars in our original proper motion sample$^4$,
we have now detected significant accelerations for three stars, S0-1, S0-2 and 
S0-4 (see Table 1); specifically, these are the sources for which
the reduced chi-squared for a quadratic fit is smaller than that for the 
linear model of their motions by more than 1.  These three stars are 
independently distinguished
in our sample, being among the fastest moving stars (v = 560 to 1350 
km/sec) and among the closest to the nominal position of Sgr A*
(r$_{1995}$ = 0.004 to 0.013 pc).  
With accelerations of 2-5 milli-arsecond/yr$^2$,
or equivalently $(3-6) \times 10^{-6}$ km/sec$^2$,
they are experiencing accelerations similar to the Earth in its orbit
about the Sun. 

Acceleration vectors, in principle, are more precise tools than
the velocity vectors for studying the central mass distribution.
Even projected onto the plane of the sky, each acceleration vector 
should be oriented in the direction of the central mass, assuming a spherically
symmetric potential.
Thus, the intersection of multiple acceleration vectors is
the location of the dark mass.
Figure 2 shows the acceleration vector's direction
for the three stars.  Within 1 $\sigma$, these vectors do indeed overlap, and
furthermore, the intersection point lies a mere 
0.05 $\pm$ 0.03 arcsec East and 0.02 $\pm$ 0.03 
arcsec South of the nominal position of Sgr A*, consistent with 
the identification of Sgr A* as the carrier of the mass.
Previously, with statistical treatments of velocities only,
the dynamical center was located to within $\pm$ 0.1 arcsec (1$\sigma$)
of Sgr A*'s position$^4$.  This velocity-based measurement is
unaffected by the increased time baseline, as its uncertainty is
dominated by the limited number of stars at a given radius.
Therefore, the accelerations 
improve the localization of our Galaxy's dynamical center 
by a factor of 3, which is critical for reliably associating any 
near-infrared source with the black hole given the complexity of the 
region.

Like the directions, the magnitudes of the acceleration vectors
also constrain the central black hole's properties.
In three dimensions, the acceleration and radius vectors 
($a_{3-D}$ and ${r_{3-D}}$, respectively)
provide
a direct measure of the enclosed mass simply by 
$ a_{3-D} = G \times M / {r_{3-D}}^2 $.   For a central potential,
the acceleration and radius vectors are co-aligned and,
with a projection angle to the plane of the sky $\theta$,  
the two dimensional projections place the following lower limit on the central
mass: $M cos^3 ( \theta ) = a_{2-D} \times {r_{2-D}^2} / G$.  
This analysis is independent of the
star's orbital parameters, although $\theta$ is in fact a lower
limit for the orbital inclination angle.
Figure 3 shows the minimum mass implied by each star's
two-dimensional acceleration as a function of projected radius,
along with dashed curves displaying how the implied mass and radius grow
with projection angle.  For each point, the uncertainty in the position
of the dynamical center dominates the minimum mass uncertainties.
Also plotted are the results 
from the statistical analysis of the velocity vectors 
measured in the plane of the sky, which imply a dark mass of 
$2.3-3.3 \times 10^6 M_{\odot}$ inside a radius of 0.015 pc$^{3,4}$. 
If, as has been assumed, this dark mass is in the form
of a single supermassive black hole, the enclosed mass should
remain level at smaller radii.  
Projection angles of 51 - 56 and 25 - 37 degrees for 
S0-1 and S0-2, respectively,  would yield the mass inferred from 
velocities and place these stars at a mere $\sim$0.008 pc
(solid blue portions of the limiting mass curves in Figure 3),  
thus increasing
the dark mass density implied by velocities by an order of magnitude
to $8 \times 10^{12} M_{\odot}/pc^3$. 
With smaller projection angles, these two stars also
allow for the enclosed mass to decrease at smaller radii,
as would occur in the presence of an extended distribution 
of dark matter surrounding a less massive black hole$^{13-17}$.
In contrast to the agreement between the mass distribution inferred from
the velocities and the accelerations for S0-1 and S0-2, 
the minimum mass implied by S0-4's acceleration is inconsistent
at least at the 1$\sigma$ level. We note, in support
of the validity of S0-4's
acceleration vector, that its orientation is consistent with the
intersection of the S0-1 and S0-2 acceleration vectors (see Fig. 2).
Nonetheless, continued monitoring of S0-4 will be important for assessing 
this possible discrepancy.
Overall, the individual magnitudes of the three acceleration vectors 
support the existence of a central black hole with mass $\sim 3 \times 10^6
M_{\odot}$.

With acceleration measurements, it is now possible to constrain the 
individual orbits.  We checked the conclusions of the 
previous studies$^{3,4}$ by assuming that the stars are bound to a 
central mass of 
$2.3 - 3.3 \times 10^6 M_{\odot}$ located within 0.03 arcsec of the nominal
position of Sgr A*.  Excellent orbital fits 
were found for S0-1 and S0-2 for the entire range of masses and for true
focii within 0.01 arcsec of the nominal position of Sgr A*, suggesting
that we now have comparable accuracy in determining the IR location of 
Sgr A* (ref. 9) and the dynamical center of our
Galaxy (see Fig. 4).  The orbital solutions for these stars have 
eccentricities ranging from 0 (circular) to 0.9 for S0-1 and 0.5 to 0.9 
for S0-2
and periods ranging from 
35 - 1,200 and 15 - 550 years, respectively, raising the possibility
of seeing a star make a complete journey around the center of the Galaxy
within the foreseeable future.
Although the fits are not yet unique, they impose a maximum
orbital distance from the black hole, or apoapse, of 0.1 pc.  
This suggests
that S0-1 and S0-2 might have formed locally.  
If these stars are indeed young$^2$, then their
small apoapse distance presents a challenge to classical star formation 
theories in light of the strong tidal forces created by
the central black hole and might require a collisional or compressional
star formation scenario$^{18,19}$. 
However, 
dynamical friction may by able to act on a short enough time scale to bring 
these stars in from a much larger distance$^{20}$.
More accurate orbital parameters are needed to fully address these problems
and others such as the distance to the Galactic Center$^{21}$.
The determinations of these parameters will be considerably improved when 
radial velocities are measured for these stars using adaptive optics 
techniques, as the current solutions based on the proper motion data alone
predict radial velocities ranging from 200 to 2000 km/sec.

{\it Acknowledgments}
This work was supported by the National Science Foundation and
the Packard Foundation.  We are grateful to James Larkin for
swapping telescope time; Ortwin Gerhard, Mike Jura, and Alycia Weinberger 
for useful input; and telescope observing assistances 
Joel Aycock, Teressa Chelminiak,  
Gary Puniwai, Chuck Sorenson, Wayne Wack, Meg Whittle, and
software/instrument specialists Al Conrad and Bob Goodrich for their help 
during the observations.  The data presented here were obtained at the W.M. 
Keck Observatory, which is operated as a scientific partnership among the 
California Institute of Technology, the University of California and the 
National Aeronautics and Space Administration.  The Observatory was made 
possible by the generous financial support of the W.M. Keck Foundation.

\noindent
Correspondence and requests for materials should be addressed
to A.G. (e-mail: ghez@astro.ucla.edu).

\pagebreak

\begin{deluxetable}{llll}
\tablenum{1}
\footnotesize
\tablecaption{Measurements for Stars with Significant Accelerations}
\tablehead{\colhead{}&
           \colhead{S0-1 (S1)} &
           \colhead{S0-2 (S2)} &
           \colhead{S0-4 (S8)} }
\startdata
Radius from Sgr A*-Magnitude (milli-pc) & 4.42 $\pm$ 0.03 & 5.83 $\pm$ 0.04 & 13.15 $\pm$ 0.04 \nl
Radius from Sgr A*-Position Angle (degrees) & 290.1 $\pm$ 0.7 & 3.1 $\pm$ 0.4 & 117.3 $\pm$ 0.2 \nl
Velocity-Magnitude (km/sec)  & 1350 $\pm$ 40 & 560 $\pm$ 20 & 990 $\pm$ 30 \nl
Velocity-Position Angle (degrees) & 168 $\pm$ 2 & 241 $\pm$ 2 & 129 $\pm$ 2 \nl
Acceleration-Magnitude (milli-arcsec/yr$^2$)& 2.4 $\pm$ 0.7 & 5.4 $\pm$ 0.3 & 3.2 $\pm$ 0.5 \nl
Acceleration-Position Angle (degrees) & 80 $\pm$ 15 & 154 $\pm$ 4  &  294 $\pm$ 7\nl
\enddata
\tablecomments{
1 - The primary nomenclature adopted here and in the text is from 
Ghez et al. (1998); in parentheses star names from Eckart \& Genzel
(1997) are also given;
2 - all quantities listed are derived from the second order 
polynomial fit of the data and are given for epoch 1995.53; 
3 - all position angles are measured East of North; 
4 - all uncertainties are 1$\sigma$ and are determined by the
jackknife resampling method$^{22}$;
5 - the radius uncertainties listed include only the relative positional 
uncertainties and not the uncertainty in the origin used 
(the nominal position of Sgr A*; see text for offset to measured dynamical
center).}
\end{deluxetable}

\clearpage

\pagebreak

\figcaption{
East-West [{\it left}] and North-South [{\it right}] 
positional offsets from the nominal
location of Sgr A* vs. time for 
S0-1[{\it top}], S0-2 [{\it middle}], \& S0-4 [{\it bottom}].
The offset range shown
is scaled to the points in each plot and therefore varies from $\sim$
0.$\tt''$04 to $\sim$ 0.$\tt''$15.
Each of these stars is located with a precision of $\sim$ 1 - 5 
milli-arcseconds
in the individual maps and the alignment of these positions,
which is carried out by minimizing the net displacement of all
stars as described in Ghez et al.$^4$, has an uncertainty of $\sim$
3 milli-arcseconds based on the half sample bootstrap resampling method$^{22}$
(the alignment uncertainty is at a minimum at the center of the map
and grows linearly with radius).  These two uncertainty terms are
added in quadrature; the results are used in the fitting process
and depicted as errorbars here.
In each plot, the solid line shows 
the second order polynomial used to derive the acceleration term.
These plots demonstrate that we are able to measure, for the first time,
accelerations of 2 to 5 milli-arcsecond/yr$^2$ (0.3-0.6 cm/s$^2$)
for stars orbiting a supermassive black hole.
}

\figcaption{
The acceleration uncertainty cones and their intersections.
The cones' edges represent the directions for which the accelerations 
deviate by 1 $\sigma$ from their best fit values and their vertices
are the time-averaged positions, 
measured relative to the nominal position of Sgr A*,
rather than the 
positions listed in Table 1.  If the accelerations
are caused by a single supermassive black hole, or even a spherically
symmetric mass distribution, these vectors should intersect at 
a common location, the center of the mass.  The 1$\sigma$ intersection point
lies 0.05 $\pm$ 0.03 arcsec East and 0.02 $\pm$ 0.03 arcsec South 
of the nominal position of Sgr A*.  The existence of an
intersection point suggests that there is indeed a common origin
for the acceleration and pinpoints the location of the black hole
to within 0.03 arcsec.}

\figcaption{
The minimum enclosed mass implied by each star's acceleration
measurement vs. projected distance from the newly determined dynamical
center position.  If the projection
between the plane of the sky and both the radius and the acceleration vectors
is $\theta$, then the true mass increases as 1/cos$^3$($\theta$) and the
true acceleration and radius vectors increase as 1/cos($\theta$).
Also plotted is the mass range implied from a statistical analysis
of the $\sim$ 100 velocity vectors that have been measured in the plane
of the sky$^{3,4}$.
}

\figcaption
{ The measured motion of S0-1, S0-2, \& S0-4 and several allowed orbital
solutions.  Only the measurements obtained in $\sim$June of each of the
5 years are shown.  S0-1 \& S0-2 are moving clockwise about
Sgr A* and S0-4 is traveling radially outward.
In the orbital analysis, two constraints are applied, a
central mass of $2.6 \times 10^6 M_{\odot}$ and a true focus located
at the position of Sgr A*.
Displayed are orbital solutions
with periods of 17, 80, 505 years for S0-2 and 63, 200, 966 years
for S0-1.
}

\pagebreak

{\centering \leavevmode
\hspace*{0.1\columnwidth} 
\epsfxsize=.80\columnwidth \epsfbox{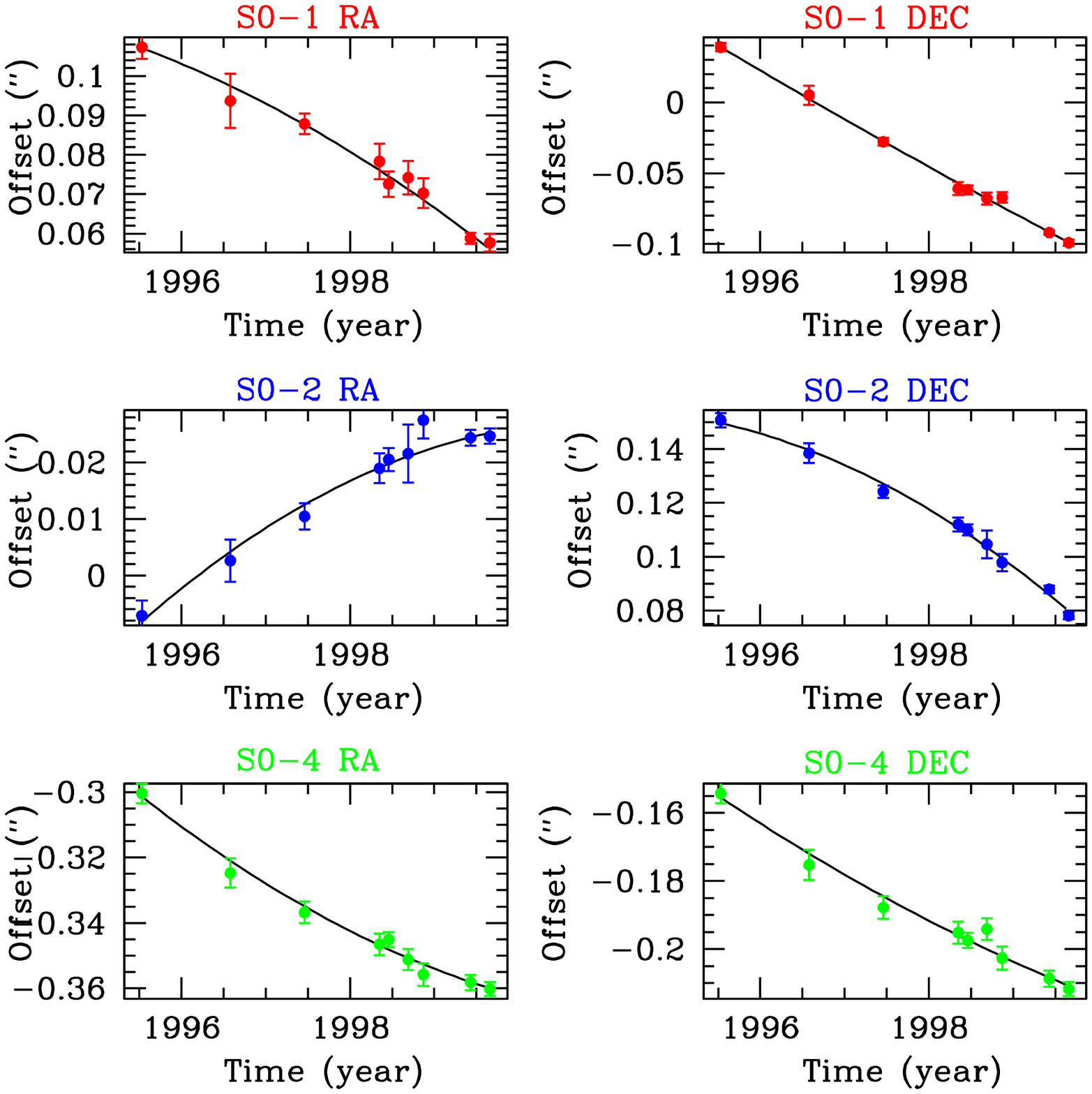}
}

\noindent
Figure 1. Ghez et al. 

{\centering \leavevmode
\hspace*{0.1\columnwidth} 
\epsfxsize=.80\columnwidth \epsfbox{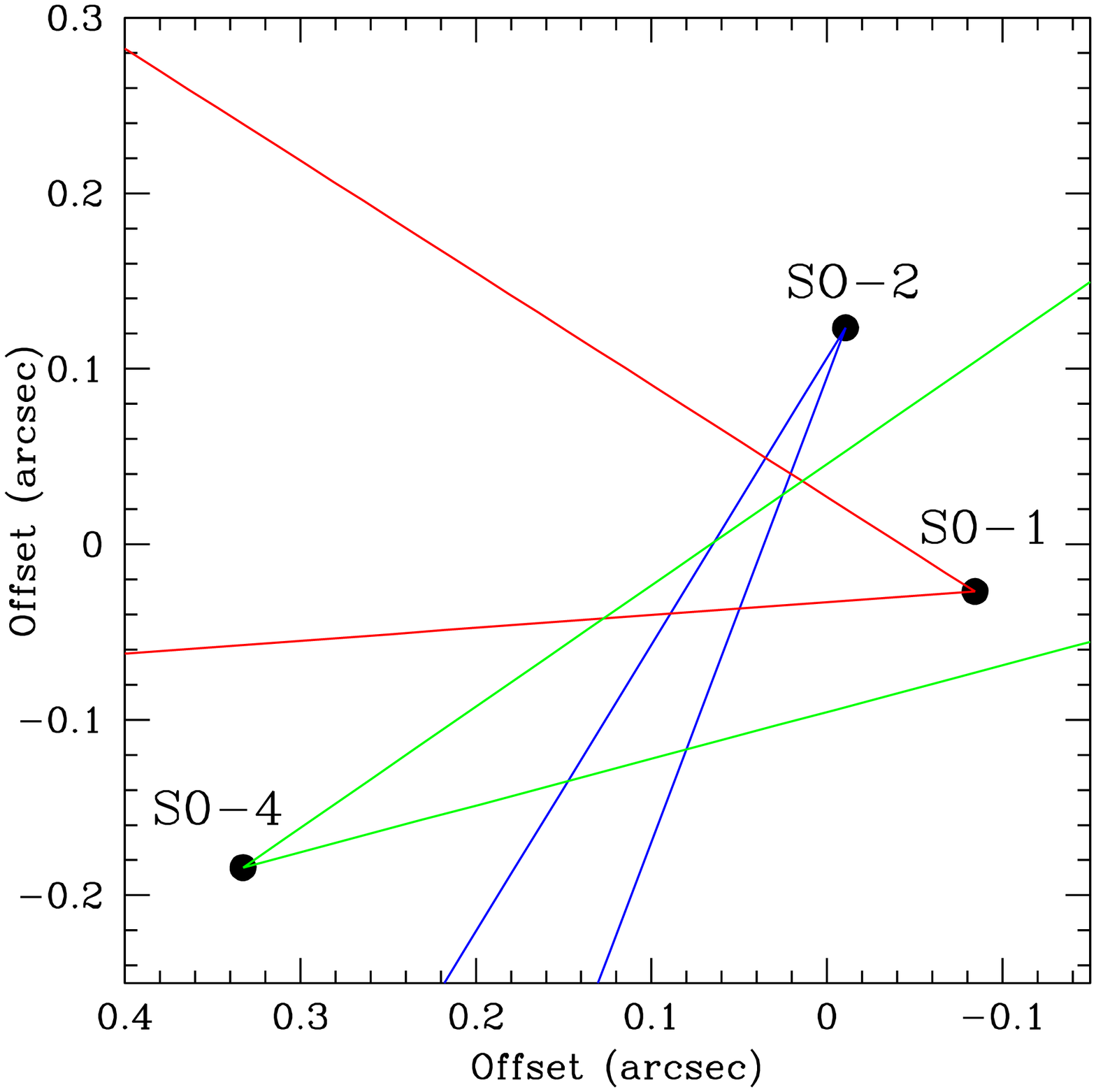}
}

\noindent
Figure 2. Ghez et al. 

{\centering \leavevmode
\hspace*{0.1\columnwidth} 
\epsfxsize=.80\columnwidth \epsfbox{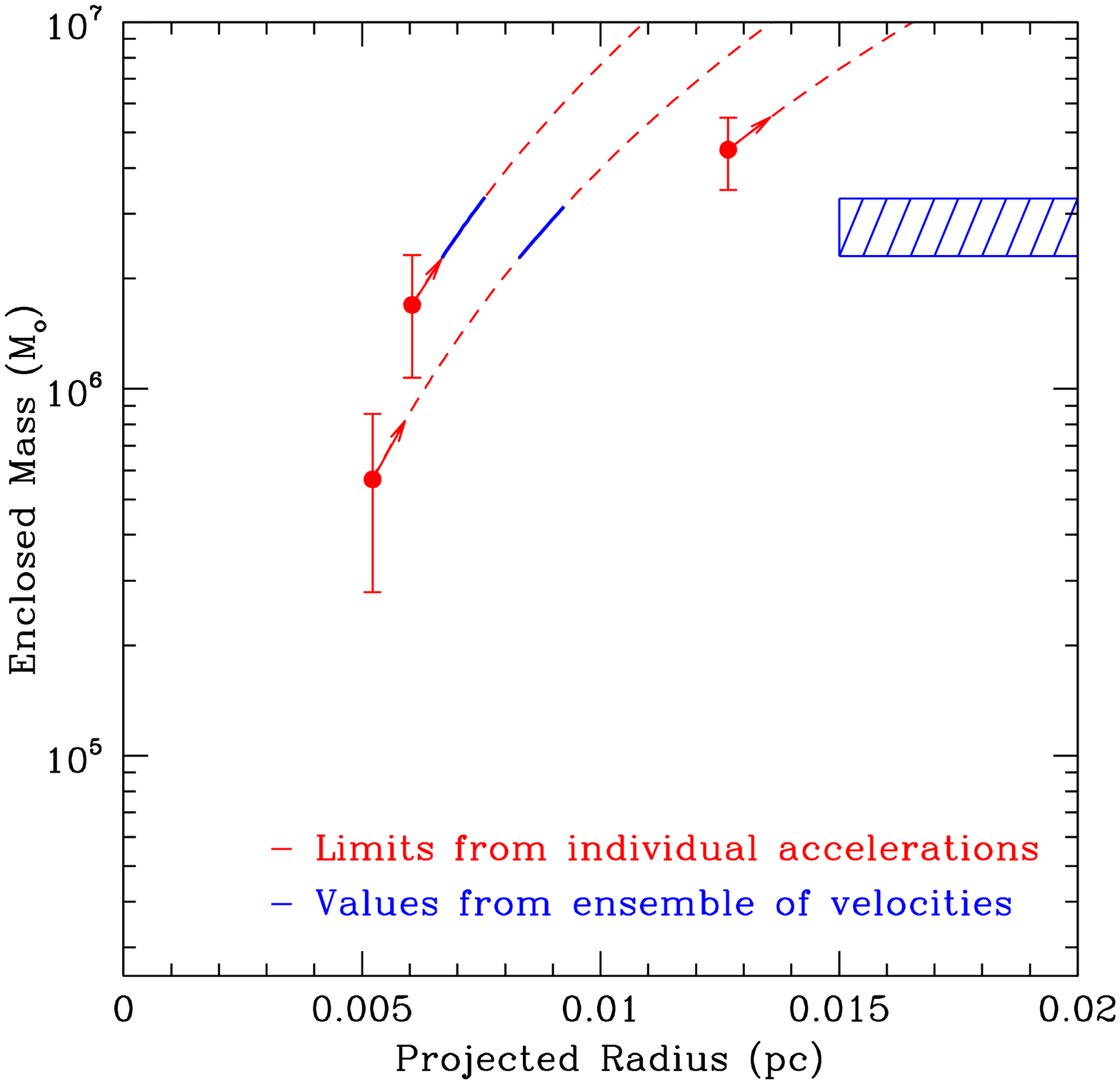}
}

\noindent
Figure 3. Ghez et al. 

{\centering \leavevmode
\hspace*{0.1\columnwidth} 
\epsfxsize=.80\columnwidth \epsfbox{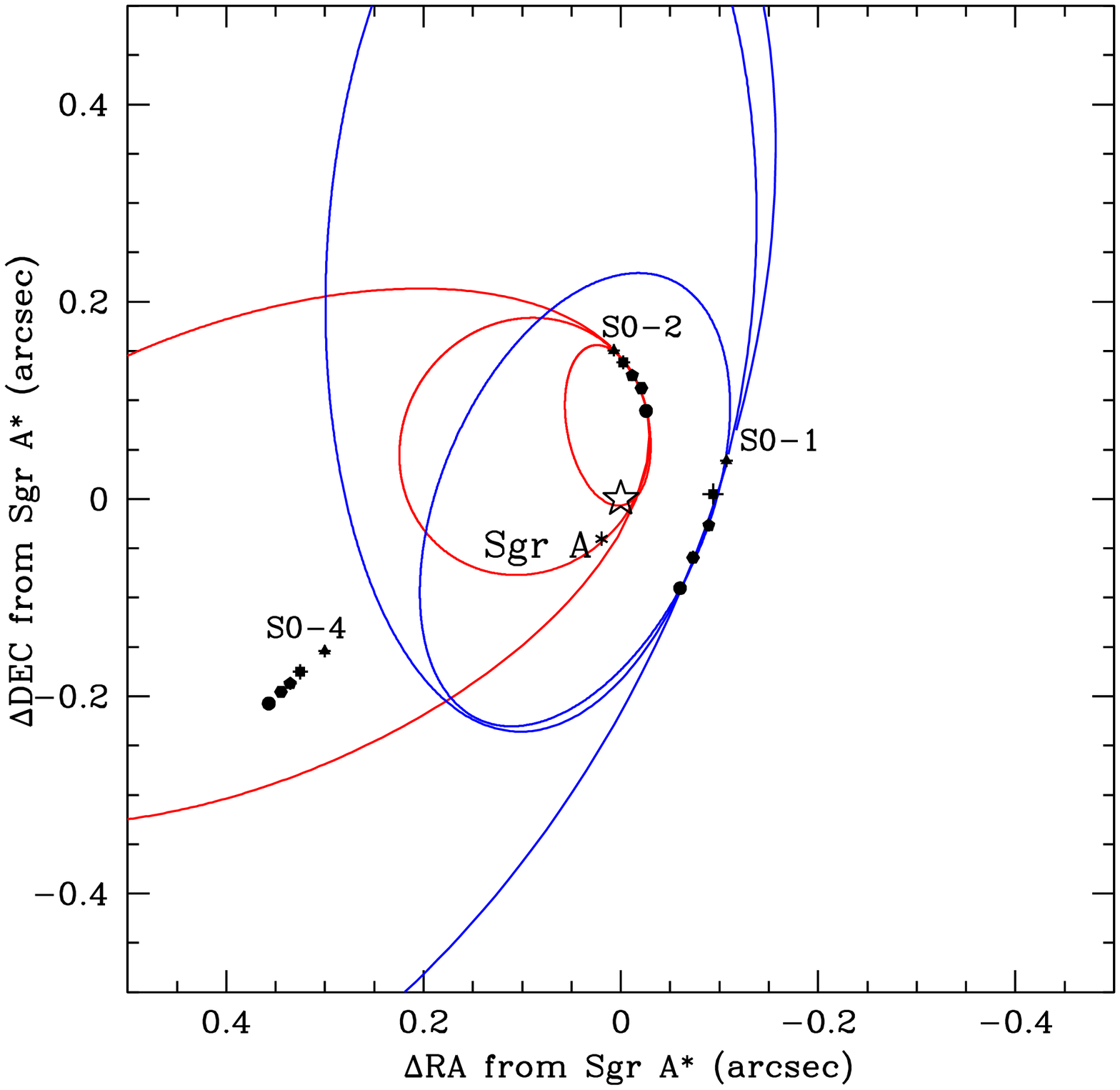}
}

\noindent
Figure 4. Ghez et al. 

\end{document}